# Electron interactions in strain-induced zero-energy flat band in twisted bilayer graphene near the magic angle

Yu Zhang[1, §], Zhe Hou[2, §], Ya-Xin Zhao[1], Zi-Han Guo[1], Yi-Wen Liu[1], Si-Yu Li[1], Ya-Ning Ren[1], Qing-Feng Sun[2,3,4,†], and Lin He[1,5,†]

[1] Center for Advanced Quantum Studies, Department of Physics, Beijing Normal University, Beijing, 100875, People's Republic of China

[2] International Center for Quantum Materials, School of Physics, Peking University, Beijing, 100871, China

[3] Collaborative Innovation Center of Quantum Matter, Beijing 100871, China

[4] Beijing Academy of Quantum Information Sciences, West Bld. #3, No. 10 Xibeiwang East Road, Haidian District, Beijing 100193, China

[5] State Key Laboratory of Functional Materials for Informatics, Shanghai Institute of Microsystem and Information Technology, Chinese Academy of Sciences, 865 Changning Road, Shanghai 200050, People's Republic of China

[§]These authors contributed equally to this work.

[†]Correspondence and requests for materials should be addressed to Qing-Feng Sun (email: sunqf@pku.edu.cn) and Lin He (e-mail: helin@bnu.edu.cn).

**In the vicinity of the magic angle in twisted bilayer graphene (TBG), the two low-energy van Hove singularities (VHSs) become exceedingly narrow[1-10] and many exotic correlated states, such as superconductivity, ferromagnetism, and topological phases, are observed[11-16]. Heterostrain, which is almost unavoidable in the TBG, can modify its single-particle band structure and lead to novel properties of the TBG that have never been considered so far. Here, we show that heterostrain in a TBG near the magic angle generates a new zero-energy flat band between the two VHSs. Doping the TBG to partially fill the zero-energy flat band, we observe a correlation-induced gap of about 10 meV that splits the flat band. By applying perpendicular magnetic fields, a large and linear response of the gap to magnetic fields is observed, attributing to the emergence of large orbital magnetic moments in the TBG when valley degeneracy of the flat band is lifted by electron-electron interactions. The orbital magnetic moment per moiré supercell is measured as about 15 $\mu_B$ in the TBG.**

In electronic flat bands, the kinetic energy of the charge carriers is quenched and electron-electron interactions may lead to the emergence of strongly correlated phases. A celebrated example is the appearance of low-energy flat bands in magic-angle twisted bilayer graphene (TBG)[1-10]. When the twist angle is finely tuned to the "magic angle", the low-energy van Hove singularities (VHSs) in the TBG become exceedingly narrow, *i.e.*, become flat bands[7-10], and many exotic correlated phases, such as superconductivity, ferromagnetism, and quantum anomalous Hall states, are observed in experiments very recently[11-16]. An alternative route to realize flat bands in 2D membrane is to introduce strain, which is demonstrated explicitly to generate flat energy bands, *i.e.*, pseudo-Landau levels[17-32], in graphene systems. However, there is no experimental evidence of many-body correlations in the strain-induced flat bands of graphene up to now. In this work, we demonstrate that heterostrain in the TBG near the magic angle can generate a new zero-energy flat band between the two low-energy VHSs and show clear evidence of strongly electron-electron correlations in the strain-induced zero-energy flat band. In our experiment, the strong many-body correlations are clearly manifested by splitting of the zero-energy flat band when it is partially filled.

In our experiments, large-area aligned graphene monolayer was grown on copper foils via a chemical vapor deposition (CVD) method and the aligned graphene monolayer was cut into two pieces to fabricate the TBG with a target twist angle $\theta$[33,34]. Then, the obtained TBG was transferred onto a $SiO_2/Si$ wafer covered by a Bernal-stacked bilayer graphene, as schematically shown in Fig. 1(a) (see Methods and Supplemental Material Figs. S1-S3 for details). The Bernal-stacked bilayer graphene can efficiently reduce electronic inhomogeneities of the TBG arising from the $SiO_2/Si$ substrate. The twist angle between the TBG and the supporting Bernal-stacked bilayer graphene is larger than 10 ° to ensure that they are electronically decoupling. By using the back-gated device and scanning tunneling microscope (STM), as shown in Fig. 1a, we can systematically study electronic properties of the TBG as a function of carrier density. Figure 1b shows a representative STM topographic image of the TBG, which exhibits a moiré superlattice with the bright spots corresponding to the AA stacking regions and the dark regions consisting of alternating AB/BA stacking regions[35-38]. The periods of the moiré pattern along the principal directions exhibit obvious anisotropic, indicating that there is a small amount of heterostrain in the TBG[35-40]. A detailed Fourier analysis of the moiré lattice reveals that the twist angle of the TBG is about 1.2 °, which is close to the magic angle (~1.1 °), and there is a slight heterostrain that consists of both a small uniaxial heterostrain $\varepsilon_{uni}^{het} = 0.38\%$ along the horizontal direction and an even smaller biaxial heterostrain $\varepsilon_{bi}^{het} = -0.05\%$错误!未找到引用源。 in the TBG (see Fig. S4 and Supplemental Material for details of analysis). Here we should point out that distorted moiré patterns are frequently observed in slightly TBG owing to the almost unavoidable heterostrain and, importantly, the moiré pattern can serve as a magnifying glass to zoom-in the heterostrain in the TBG[35-40].

Figure 1c shows two typical scanning tunneling spectroscopy (STS), i.e., *dI/dV*, spectra measured at the AA and AB regions of the TBG. Two notable experimental features can be observed according to the results shown in Fig. 1c. The first is that the spectra feature three sharp peaks (labelled as $E_{-VHS}$, $E_{+VHS}$, and $E_0$), which are much more intense in the AA regions. To explore the nature of these peaks, we carried out the STS maps, which can directly reflect the spatial distributions of the local density of states (LDOS), at the energies of these peaks (Fig. S5). Our measurement demonstrates that the electronic states of the three peaks are mainly localized in the AA regions of the moiré pattern. As an example, Figure 1d shows the STS map recorded at the energy of $E_0$, which exhibits the same structure as the moiré pattern in the TBG. Usually, there are two low-energy VHSs flanking the charge neutrality point in slightly TBG and the electronic states of the two VHSs are mainly localized in the AA regions of the moiré pattern[1-7,35-38]. Therefore, the two peaks flanking the $E_0$ are attributed to the two VHSs of the TBG. The $E_0$ peak around the charge neutrality point is attributed to the flat band that is generated by the heterostrain in the TBG. Very recently, it was demonstrated explicitly that the heterostrain can introduce a flat band between the two VHSs of a 1.25 °TBG[39]. To further confirm the origin of the $E_0$ peak, we carry out tight-binding calculation by taking into account the heterostrain of the TBG (Fig. 2a) and Fig. 2b shows the calculated LDOS of the TBG (see Supplemental Material for details). There are three low-energy peaks, which are mainly localized in the AA regions, in the LDOS of the TBG and the heterostrain in the TBG generates a zero-energy flat band between the two VHSs. Obviously, our theoretical results reproduce the main feature of our experimental observation. In our experiment, the energy separation between the two VHSs, ~ 120 meV, of the TBG is much larger than that obtained in theory ~ 50 meV. The giant enhanced energy separation between the VHSs is attributed to the strong *e-e* interactions when the chemical potential lies between the two VHSs. Very recently, similar behavior has been observed in the magic-angle TBG and is treated as evidence for the strong many-body correlations in the magic-angle TBG[35-38].

The second notable feature observed in the spectra of Fig. 1c is the appearance of several weak peaks at high energies, as marked by the red arrows (see Figs. S5 and S6 for more experimental data). These peaks are attributed to strain-induced pseudo-Landau levels in the TBG. The strain generates pseudo-magnetic fields that confine the massive Dirac electrons in the TBG into circularly localized pseudo-Landau levels, as demonstrated very recently in low-angle TBG[40,41]. Our theoretical calculation also demonstrates that the heterostrain can generate high-energy peaks in the DOS of the TBG, as shown in Fig. 2b. The energy spacing between the pseudo-Landau levels obtained in theory is slightly larger than that measured in experiment, which may arise from a relatively larger strain of the TBG in our simulation.

In the magic-angle TBG, the Coulomb interactions can greatly exceed the kinetic energy of the charge carriers and lead to various strongly correlated phases when the Fermi energy lies within the flat bands[11-16,35-38]. In this work, the kinetic energy of electrons in the strain-induced flat band can be estimated by the band width as $W_0 = 16.8\ meV$, which is much smaller than the on-site Coulomb energy of each site $U \sim e^2/\varepsilon L \sim 27\ meV$ ( $e$ is the electron charge, $\varepsilon \sim 4.4\ \varepsilon_0$ with $\varepsilon_0$ the vacuum dielectric constant, and $L \sim 12\ nm$ is the moiré period) in the studied TBG. Therefore, bringing the Fermi energy within the strain-induced flat band is also expected to lead to various strongly correlated phases. To explore this, we measure the properties of the TBG as a function of electron density in the back-gated devices. Figure 3a shows representative d*I*/d*V* spectra measured on the AA regions of the moiré pattern in the TBG as a function of back-gate voltage $V_g$. When 0 V < $V_g$ < 12 V, the strain-induced flat band of $E_0$ is fully unoccupied and the tunneling spectrum is almost independent of the back-gate voltage. However, when we continuously increase the back-gate voltage ($V_g$ > 12 V) and the strain-induced flat band becomes partially filled, a notable new feature is observed in the spectra: the strain-induced flat band splits into two peaks (see Fig. S7 for details of the splitting). Figure 3b summarizes the splitting of the flat band as a function of the back-gate voltage. Simultaneously, we also observe a slight broadening of both the fully occupied VHS and the completely empty VHS when the strain-induced flat band is partially filled, as shown in Fig. 3c and 3d. The full width at half maximum (FWHM) of the two VHSs increases simultaneously with the splitting of the strain-induced flat band. Such a phenomenon is unexpected since that there is no reason to expect a broadening of the fully occupied (or completely empty) bands when the occupation of the other bands is changed. The above two features, including the splitting of the partial-filled flat bands and the broadening of the fully occupied (or completely empty) VHSs, observed in this work are quite similar as the strong many-body correlations observed in the magic-angle TBG[35-38]. In the magic-angle TBG, when one of the VHSs (or flat bands) is partially filled, both a correlation-induced gap of about 10 meV in the partial-filled flat band and a broadening of the other flat band (either fully occupied or completely empty) are also observed. This indicates that the *e-e* interactions are also very important in the strain-induced flat band and, therefore, it is possible to realize interesting correlated phenomena in the strain-induced flat band.

Although the correlation-induced gap is clearly observed in this work and in the magic-angle TBG when the flat band is partially filled, the nature of the gap observed in the STM measurements[35-38] is still unclear up to now. A careful examination of the back-gate-dependent spectra, as shown in Fig. 3a, shows that the intensities of the split peaks on either side of the Fermi level are about half the intensity of the un-split flat band, which indicates that either the spin or the valley degeneracy of the flat band is lifted by the *e-e* interactions. To further explore the nature of the splitting at partial filling of the strain-induced flat band, we measure high-resolution STS spectra of the

TBG in different doping levels as a function of perpendicular magnetic fields, as summarized in Fig. 4. The tunneling spectra are almost independent of the magnetic fields when the strain-induced flat band is completely empty, as shown in Fig. 4a. However, when the strain-induced flat band is partially filled, the correlation-induced splitting increases quickly with increasing the magnetic fields, as shown in Fig. 4b (see Fig. S8 for detailed analysis of the splitting in different magnetic fields). Figure 4c shows the splitting $\Delta E$ as a function of magnetic field, which exhibits a linear scaling. Derived from the slope of the linear fit of a Zeeman-like splitting $\Delta E = g_{eff} \mu_B B$, we obtain an effective $g$ factor as about 30. The obtained giant effective $g$ factor is much larger than that of the Zeeman splitting of spin (the effective $g$ factor of spin is about 2) and is almost the same as that of the valley splitting in graphene monolayer when the zero Landau level is half filled[42,43]. Such a result indicates that the correlation-induced splitting lifts the valley degeneracy of the strain-induced flat band. Then, the observed large and linear response of the splitting to magnetic fields is attributed to coupling to large orbital magnetic moments in the TBG. Theoretically, it was predicted that large orbital magnetic moments can be generated by circulating current loops in the moiré patterns of slightly TBG when the valley degeneracy of the flat band is lifted by the *e-e* interactions and the $C_{2z}$ symmetry of the TBG is broken by the substrate[44-51]. In our experiment, the *e-e* interactions lift the valley degeneracy of the strain-induced flat band and the supporting Bernal bilayer graphene can break the $C_{2z}$ symmetry of the TBG. Therefore, each valley would be associated with non-vanishing Berry curvature, resulting in the emergence of the large orbital magnetic moments in the studied system. In our experiment, the spatial distribution of the strain-induced flat band exhibits the same structure as the moiré pattern in the TBG. Therefore, the orbital magnetic moment in each moiré of the TBG is estimated as about 15 $\mu_B$, according to the obtained slope in Fig. 4c. This is consistent with the orbital magnetic moment in each moiré of the magic-angle TBG obtained in theory[44-51].

In summary, our results introduce a new platform to realize various strongly correlated phases, not limited in the magic-angle TBG, which requires fine-tuning of the twist angle. We demonstrate explicitly that the strong electron-electron correlations dominate the electronic properties in the strain-induced zero-energy flat band in the TBG near the magic angle. The *e-e* interactions lift the valley degeneracy of the strain-induced flat band and lead to the emergence of large orbital magnetic moment in each moiré of the TBG when the flat band is partially filled. It indicates that we can explore magnetism that is purely orbital in the TBG.

**Acknowledgements**

This work was supported by the National Natural Science Foundation of China (Grant Nos. 11974050, 11674029) and National Key R and D Program of China (Grant No. 2017YFA0303301). L.H. also acknowledges support from the National Program for Support of Top-notch Young Professionals, support from "the Fundamental Research Funds for the Central Universities", and support from "Chang Jiang Scholars Program".

**Author contributions**

Y.Z. and Y.X.Z. synthesized the samples and performed the STM experiments. Z. H. and Q.F.S. performed the theoretical calculations. Y.Z. analyzed the data. L.H. conceived and provided advice on the experiment, analysis, and the theoretical calculation. L.H. and Y.Z. wrote the paper. All authors participated in the data discussion.

**Competing financial interests**

The authors declare no competing financial interests.

**Figures**

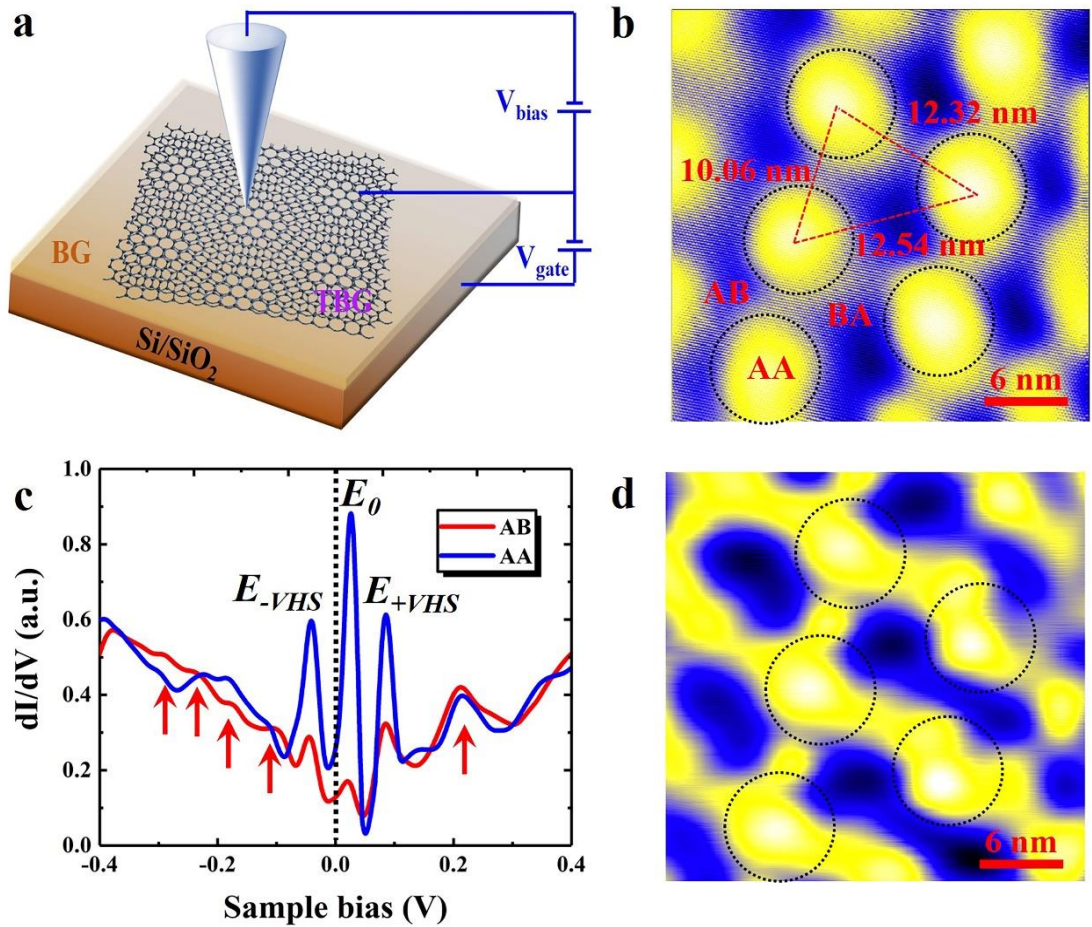

**Figure 1. STM characterization of a TBG near the magic angle under heterostrain.**
**(a)** Schematic of the STM setup on the TBG devices. **(b)** A $30 \times 30$ nm$^2$ STM topographic image showing the moiré superlattice with a twist angle $\theta \sim 1.2°$ and heterostrain $\varepsilon \sim 0.38\%$ ($V_b = 300\ mV, I = 0.2\ nA$). **(c)** Typical STS spectra recorded at AA and AB regions ($V_b = 300\ mV, I = 0.2\ nA$). Peaks of $E_0$ and $E_{\pm VHS}$ are marked in the figure. The black dashed line is the energy of the Fermi level. **(d)** A typical STS map recorded at the energy of $E_0$. The black dotted circles mark the period and circular symmetry of the AA regions of the moiré patterns in the TBG.

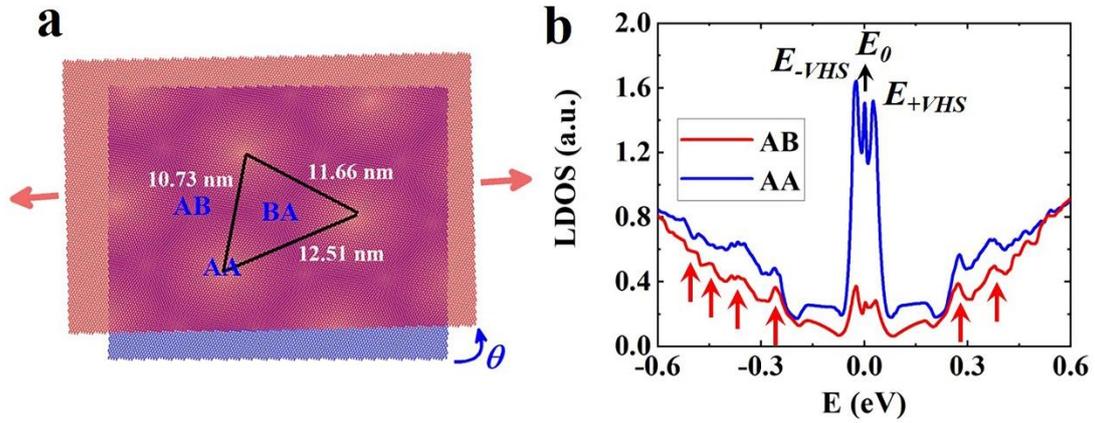

**Figure 2. Theoretical calculations of the TBG near the magic angle under heterostrain.** (**a**) Schematic of the TBG with the twist angle $\theta$ and a uniaxial strain applied to the above layer only. The calculated moiré periods are given in the figure. (**b**) The LDOS in the AA and AB stacked regions obtained from the tight-binding model. Peaks of $E_0$ and $E_{\pm VHS}$ are marked in the figure. The strain-induced high energy peaks, i.e., the pseudo-Landau levels of massive Dirac fermions, are marked by the arrows.

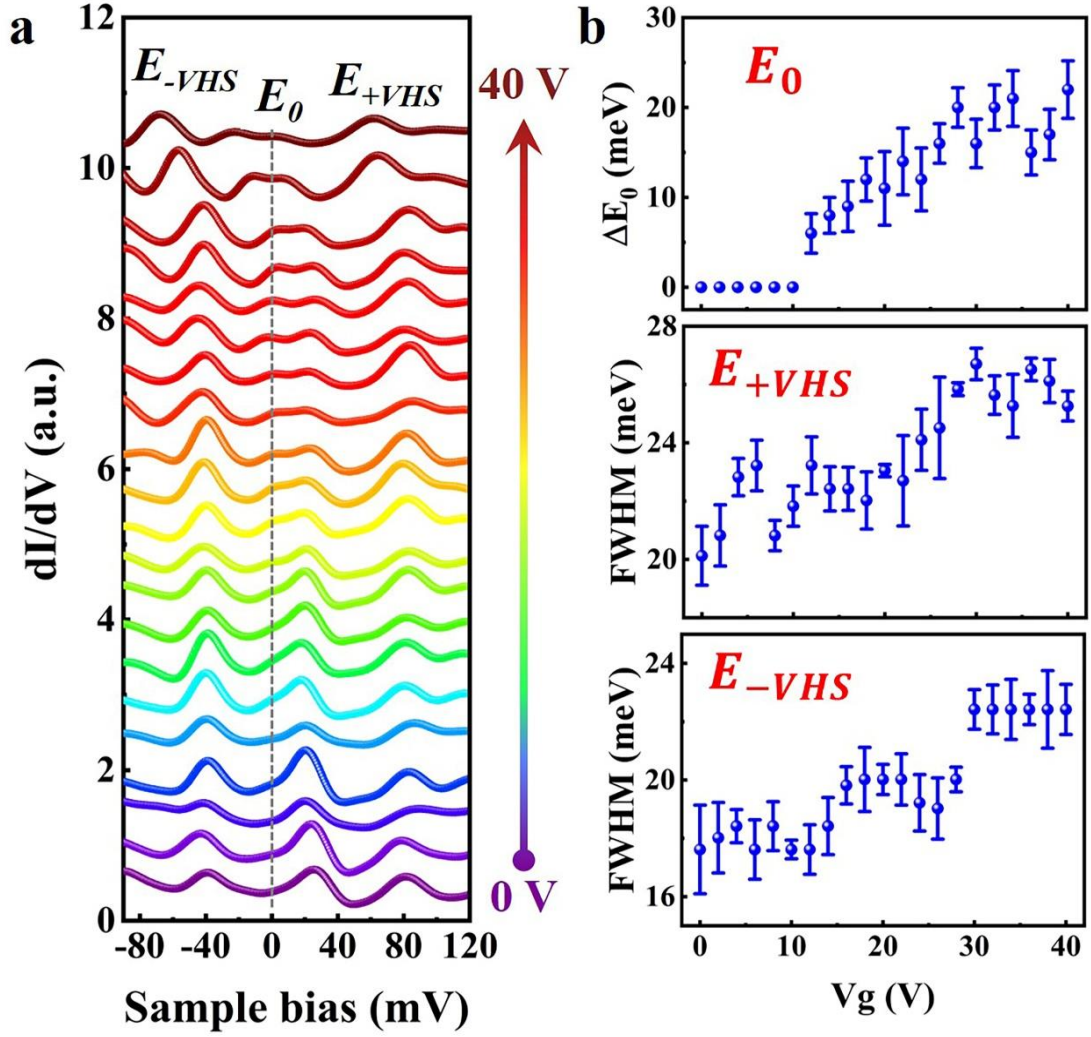

**Figure 3. High-resolution STS spectra of the TBG with different back-gate voltages. (a)** STS spectra around the charge neutrality point recorded in the center of the AA region of the TBG with different back-gate voltages. When the back-gate voltage is 0 V, the strain-induced flat band is empty. With increasing the back-gate voltage to 40 V, the flat band is almost half filled and splits into two peaks flanking the Fermi level. The gray dashed line is the energy of Fermi level, and all the curves are offset for clarity. **(b)** Up panel: the energy separations of the splitting in the strain-induced flat band as a function of $V_g$. Middle and bottom panels: The FWHM of $E_{+VHS}$ and $E_{-VHS}$ as a function of $V_g$.

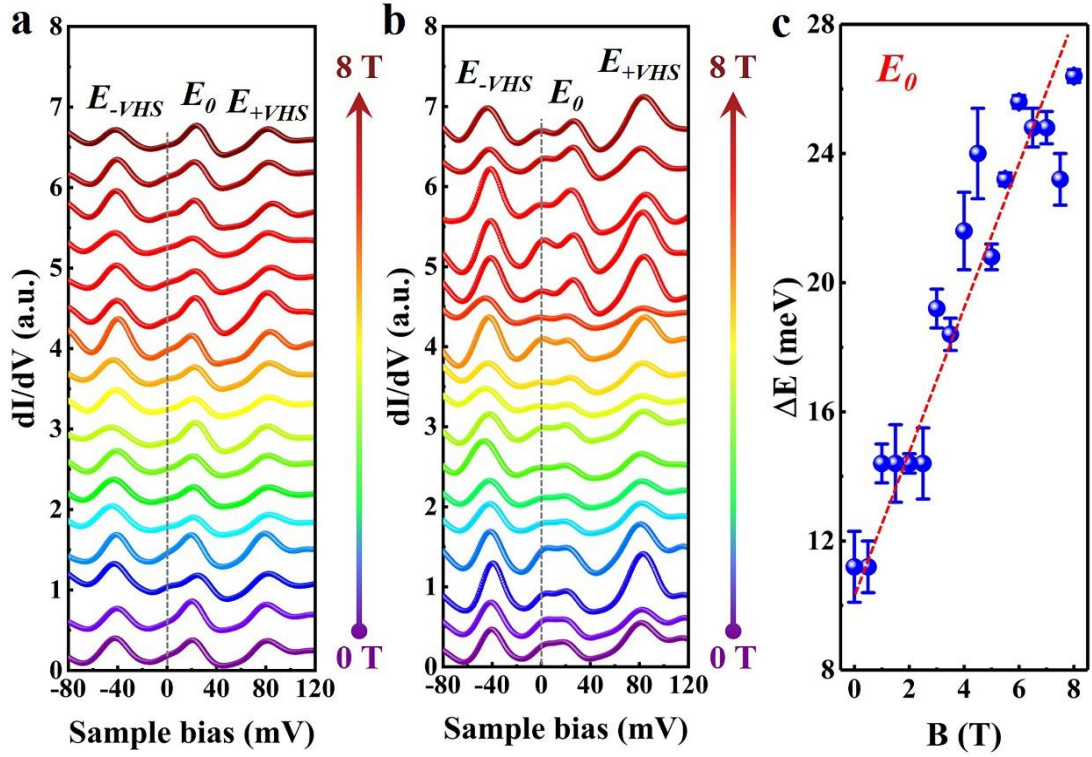

**Figure 4. High-resolution STS spectra of the TBG in different magnetic fields. (a)** Typical STS spectra measured in different magnetic fields when the strain-induced flat band is empty. **(b)** High-resolution STS spectra measured in different magnetic fields when the strain-induced flat band is partially filled. The gray dashed line is the energy of the Fermi level, and all the curves are offset for clarity. **(c)** Energy separations of the two split peaks as a function of magnetic fields. The red dashed line is the linear fit.